\documentclass[a4paper,12pt]{article}

\begin{document}

\begin{center}
\Large{\bf Variational principle
and phase space measure in non-canonical coordinates }
\end{center}
    
\begin{center}
Alessandro Sergi\footnote{ (E-mail: asergi@unime.it)}\\
Dipartimento di Fisica, Sezione Fisica Teorica,\\
Universit\'a degli Studi di Messina,\\
Contrada Papardo C.P. 50-98166 Messina, Italy
\end{center}

\begin{center}
    {\bf Abstract}\\
\end{center}
{\small Non-canonical equations of motion
are derived from a variational principle
written in symplectic form.
The invariant measure of phase space 
and the covariant expression for the entropy 
are derived from non-canonical transformations of coordinates.
This shows that the geometry of non-canonical phase space
is non trivial even if dynamics has no compressibility. 
}

\section{Introduction}

Theoretical formalisms very often  use non-canonical equations of motion.
For example, the equations for Eulerian variables,
that describe ideal continuous media, are in general
non-canonical~\cite{morrison}.
Non-canonical phase space flows can be derived from Hamiltonian dynamics
by means of non-canonical transformations of phase space coordinates
(i.e. transformations with Jacobian not equal to one)
while non-Hamiltonian dynamics cannot be derived using only
transformations of phase space coordinates.
However, non-canonical dynamics has a certain likeness with
energy-conserving non-Hamiltonian dynamics~\cite{b1}
(this latter is commonly used in molecular dynamics simulations).
For this reason, non-canonical systems can be used
to improve our understanding of non-Hamiltonian systems
with a conserved energy~\cite{b1}.
In the following, the comparison between non-Hamiltonian
and non-canonical system will be exploited
to clarify some issues regarding phase space measure.
It will be shown that the invariant measure of non-canonical
phase space can be derived from coordinate transformations
without the need to consider dynamical properties,
such as phase space compressibility.
The situation is different in the non-Hamiltonian case,
where one has to resort to arguments associated
with time evolution in phase space~\cite{b1,tuckerman}.
Coordinate transformations will be also used
to obtain the covariant form of the entropy functional
and a variational principle for non-canonical equations of motion,
arising from a symplectic form of the action.

This paper is organized as follows:
in Section~\ref{sec:h} the action is written
by means of the symplectic matrix and
Hamiltonian equations in canonical form
are derived from the variational principle.
In Section~\ref{sec:nc} coordinate transformations
will be applied to the symplectic form of the action
in order to obtain non-canonical equations of motion
from the variational procedure.
In Section~\ref{sec:nh} it is shown how a certain class
of non-Hamiltonian equations may be derived by introducing
a coordinate-dependent scaling of time.
Phase space invariant measure and 
the covariant form of the entropy functional are obtained
from coordinate transformations in Section~\ref{sec:measure}.
Finally, conclusions are given in Section~\ref{sec:conclusion}.

\section{Hamiltonian Canonical Equations in Symplectic Form}\label{sec:h}

Consider a general time-independent Hamiltonian ${\cal H}$.
Points in a $2N$-dimensional phase space will be indicated by $x=(q,p)$.
Consider also the symplectic matrix~\cite{goldstein,mccauley}
\begin{equation}
{\bf B}^c=
\left[ \begin{array}{cc}
0 & 1 \\ -1 & 0
\end{array} \right]\;,
\end{equation}
with the following properties
\begin{eqnarray}
\left[{\bf B}^c\right]^{T}&=&-{\bf B}^c\;, \\
\left[{\bf B}^c\right]^{-1}&=&-{\bf B}^c\;.
\end{eqnarray}
Canonical Hamiltonian equation of motion are written 
in symplectic form~\cite{goldstein,mccauley} as
\begin{equation}
\dot{x}_{i}= B^c_{i j}\frac{\partial {\cal H}}{\partial x_{j}}
\;.  \label{canham}
\end{equation}
In the above equation and in the following ones, Einstein's convention
of summing over repeated indices must be understood.
The free indices assume all the values within their range of variation,
which depends on the dimension of phase space.
Usually, one first derives Hamiltonian equations and then shows
that they can be written in symplectic form~~\cite{goldstein,mccauley}. 
It can be shown
that Eq.s~(\ref{canham}) can be obtained directly from the variational principle
$\delta{\cal A}=\int~dt\left[p\dot{q}-{\cal H}\right]=0$, where $\cal A$
is the action.
To this end, one starts from the action
\begin{eqnarray}
{\cal A}&=&\int dt \left[ p\dot{q}-{\cal H}  \right] 
=\int~dt \left[  \frac{1}{2}\left(p\dot{q}-q\dot{p}\right)-{\cal H}\right] \;,
\end{eqnarray}
where the last equality is obtained by means of an integration by parts.
Noticing that
\begin{equation}
p\dot{q}-q\dot{p}=
\left[ \begin{array}{cc} \dot{q} & \dot{p} \end{array} \right]
\left[ \begin{array}{cc} 0 & 1 \\ -1 & 0 \end{array} \right]
\left[ \begin{array}{c} q \\ p \end{array} \right]\;,
\end{equation}
the action can be expressed by means of the symplectic matrix
${\bf B}^c$ as
\begin{equation}
{\cal A}=\int dt \left[\frac{1}{2} \dot{x}_{i}B^c_{i j}x_{j}-
{\cal H}  \right]\;. \label{symact}
\end{equation}
Applying the variational procedure one gets
\begin{eqnarray}
\delta {\cal A} &=&\int dt \left[ \frac{1}{2}\delta \dot{x}_{i}B^c_{i j}x_{j}+\frac{1}{2}\dot{x}_{i}B^c_{i j}\delta x_{j}
-\frac{\partial {\cal H} }{\partial x_{j}}\delta x_{j} \right] 
\nonumber \\
&=&\int dt \delta x_{j} \left[\dot{x}_{i}B^c_{i j}
-\frac{\partial {\cal H} }{\partial x_{j}} \right] =0 \;.
\end{eqnarray}
Finally, using $B^c_{i j}B^c_{j k}=\delta_{ik}$, 
the equations of motion~(\ref{canham}) are obtained.

\section{Non-canonical equations of motion}\label{sec:nc}

Consider a set of phase space coordinates $z=z(x)$
obtained using a transformation with
Jacobian ${\cal J}=|\frac{\partial x}{\partial z}   | \neq 1$.
Such coordinates are non-canonical~\cite{morrison,goldstein,mccauley}.
Under non-canonical transformations,
the Hamiltonian transforms as a scalar according to
${\cal H}(x(z))={\cal H}'(z)$ and Eq.s~(\ref{canham}) become:
\begin{eqnarray}
\dot{z}_{n}&=&B_{n m}\frac{\partial {\cal H}'}{\partial z_{m}}\;,
\label{ncanham}
\end{eqnarray}
where the antisymmetric matrix 
\begin{equation}
B_{n m}=\frac{\partial z_{n}}{\partial x_{i}}B^c_{i j}\frac{\partial z_{m}}
{\partial x_{j}}
\label{bnm}
\end{equation}
has been introduced.
Equations~(\ref{ncanham}) are referred as non-canonical
 Hamiltonian equations~\cite{morrison,mccauley}.
They might introduce a compressibility in phase space
\begin{equation}
\kappa=\frac{\partial\dot{z}_n}{\partial z_n}=
\frac{\partial B_{n m}}{\partial z_n}\frac{\partial {\cal H}}{\partial z_{m}}
\;,
\end{equation}
but one can also have non-canonical equations 
with zero compressibility~\cite{morrison}.
In the following, it will be shown that  Eq.s~(\ref{ncanham}) can be derived
from a variational principle.
To this end, one applies a non-canonical transformation of coordinates
to the symplectic expression of the action, Eq.~(\ref{symact}):
\begin{equation}
{\cal A}=\int dt \left[\frac{1}{2} 
\frac{\partial x_{i}}{\partial z_{m}}\dot{z}_{m}
B^c_{i j}x_{j}(z)-{\cal H}'(z)  \right] \;.
\label{nsymact}
\end{equation}
On the action expressed by Eq.~(\ref{nsymact}) one can perform
a variation on non-canonical coordinates and integrate
by parts when needed. Thus, one obtains 
\begin{eqnarray}
\delta {\cal A} &=&
\int dt \Big[\frac{1}{2}
 \frac{\partial^{2} x_{i}}{\partial z_{m}z_{k}}\delta z_{k}
\dot{z}_{m}B^c_{i j}x_{j}(z)
-\frac{1}{2}\frac{d}{dt} 
\left(\frac{\partial x_{i}}{\partial z_{m}}
B^c_{i j}x_{j}(z)\right) \delta z_{m}
\nonumber \\
&+&\frac{1}{2} \frac{\partial x_{i}}{\partial z_{m}}\dot{z}_{m}
B^c_{i j}\frac{\partial x_{j}}{\partial z_{k}}
\delta z_{k}
-\frac{\partial {\cal H}'}{\partial z_{k}}\delta z_{k}  \Big] \;.
\label{eq:delta_a_1}
\end{eqnarray}
Considering now the term
\begin{eqnarray}
-\frac{1}{2}\frac{d}{dt} \left(\frac{\partial x_{i}}{\partial z_{m}}
B^c_{i j}x_{j}(z)\right) \delta z_{m}
&=& -\frac{1}{2}\frac{d}{dt} \left(\frac{\partial x_{i}}{\partial z_{k}}
B^c_{i j}x_{j}(z)\right) \delta z_{k}
\nonumber \\ &=&
-\frac{1}{2} \left(\frac{\partial^{2} x_{i}}{\partial z_{m}\partial z_{k}}
\dot{z}_{m} B^c_{i j}x_{j}(z)\right. \nonumber\\
&+&\left.\frac{\partial x_{i}}{\partial z_{k}}
B^c_{i j}\frac{\partial x_{j}}{\partial z_{m}}\dot{z}_{m}
\right) \delta z_{k} \;,
\end{eqnarray}
and substituting into Eq.~(\ref{eq:delta_a_1}) one finds
\begin{eqnarray}
\delta {\cal A} &=&
\int dt \delta z_{k}\Big[ 
\frac{\partial x_{i}}{\partial z_{m}}B^c_{i j}
\frac{\partial x_{j}}{\partial z_{k}}\dot{z}_{m}
-\frac{\partial {\cal H}'}{\partial z_{k}}  \Big] \;,
\end{eqnarray}
which leads to equations in the form
\begin{eqnarray}
\dot{z}_m\frac{\partial x_i}{\partial z_m}B_{i j}^c
\frac{\partial x_j}{\partial z_k}
&=&\frac{\partial {\cal H}'}{\partial z_k}\;.
\end{eqnarray}
By inverting the matrix 
\begin{equation}
A_{m k}=\frac{\partial x_i}{\partial z_m}B_{i j}^c
\frac{\partial x_j}{\partial z_k} \;,
\end{equation}
one finds the equations already given
in~(\ref{ncanham}) with $B_{s k}$ given by Eq.~(\ref{bnm}).
It is worth to remark that Eq.s~(\ref{ncanham}) are still
Hamiltonian although expressed in non-canonical form~\cite{morrison,mccauley}.
As a matter of fact, a generalized bracket satisfying the
Jacobi relation can be introduced~\cite{morrison,b1}.
Considering the matrix $B_{s k}$ defined in Eq.~(\ref{bnm}),
the Jacobi relation leads to 
\begin{equation}
B_{in}\frac{\partial B_{jk}}{\partial x_n}
+B_{kn}\frac{\partial B_{ij}}{\partial x_n}
+B_{jn}\frac{\partial B_{ki}}{\partial x_n} =0\;.
\label{jacobi}
\end{equation}

\section{Non-Hamiltonian equations of motion}\label{sec:nh}

Non-Hamiltonian equations of motion with a conserved
energy~\cite{b1} may be defined using the structure of Eq.s~(\ref{ncanham}).
To this end, one uses an antisymmetric matrix
${\bf B}^{NH}$ which is no longer defined by Eq.~(\ref{bnm})
and does not satisfies the Jacobi relation as given by Eq.~(\ref{jacobi}).
Originally, Nos\'e~\cite{nose} showed that a certain class
of energy conserving non-Hamiltonian equations
can be derived by combining non-canonical transformation
of phase space coordinates and position-dependent scalings of time.
Here it is shown how this can be achieved when starting from
the variational principle in symplectic form.
To this end, one considers the expression obtained
by performing the variation of the action using non-canonical
variables
\begin{equation}
\delta {\cal A}=\int~dt\delta z_k\left(
\dot{z}_m\frac{\partial x_i}{\partial z_m}B_{i j}^c
\frac{\partial x_j}{\partial z_k}
-\frac{\partial {\cal H}'}{\partial z_k}
\right)\;.
\end{equation}
A position-dependent scaling of time is then applied
\begin{equation}
dt=\phi(z)d\tau\;.
\end{equation}
In general, the above time scaling will depend on the path in phase space.
Then one obtains
\begin{equation}
\dot{z}_s=B_{sk}^{NH}
\frac{\partial {\cal H}'}{\partial z_k}\;,
\label{nheq}
\end{equation}
with
\begin{equation}
B_{sk}^{NH}=
\phi^{-1}(z)\frac{\partial z_s}{\partial x_n}
B_{n l}^c \frac{\partial z_k}{\partial x_l}\;.
\label{nhb}
\end{equation}
Equations~(\ref{nheq}), 
are non-Hamiltonian since, in general, the Jacobi relation,
Eq.~(\ref{jacobi}), will not be satisfied
because of the new matrix defined in Eq.~(\ref{nhb}).
Phase space compressibility for Eq.~(\ref{nheq}) is defined as
\begin{eqnarray}
\kappa^{NH}&=&
-\phi^{-2}(z)
\frac{\partial \phi}{\partial z_s}B_{sk}\frac{\partial {\cal H}'}{\partial z_k}
+\phi^{-1}(z)
\frac{\partial B_{sk}}{\partial z_s}\frac{\partial {\cal H}'}{\partial z_k}\;.
\end{eqnarray}
The first term on the right hand side is proportional to the time derivative
of $\phi$ considering  its evolution under the non-canonical flow
(instead of the non-Hamiltonian one).
The second term is proportional to the compressibility arising from the
non-canonical transformation of coordinates.
This two terms are combined with the scaling function $\phi$
to give the non-Hamiltonian compressibility $\kappa^{NH}$.
If both terms are zero, the non-Hamiltonian flow will
have zero compressibility.

\section{Invariant Measure}\label{sec:measure}
 
Using canonical coordinates, averages in phase space are defined as~\cite{balescu}
\begin{equation}
\langle a\rangle=\int dx f(x) a(x) \;,
\end{equation}
where $f(x)$ is the normalized distribution function
and $a$ is an arbitrary phase space observable.
Considering a transformation to non-canonical coordinates,
one obtains 
\begin{equation}
\int dx f(x)\rightarrow \int dz\Bigg|\frac{\partial x}{\partial z}\Bigg|
f'(z)=\int dz \rho(z)\;,
\end{equation}
where the Jacobian
${\cal J}=\partial x/\partial z\neq 1$ is used to define the measure
of non-canonical phase space.
The Jacobian can be absorbed into a properly
defined distribution function
\begin{equation}
\rho(z)=\Bigg|\frac{\partial x}{\partial z}\Bigg|f'(z)\;,
\label{rhoz}
\end{equation}
which can be shown to obey a generalized Liouville
equation~\cite{b1}.
As it will be shown in the following, this is particularly important
when defining the entropy.

It is now easy to show that the measure in phase space,
introduced by the change of coordinates,
is also invariant in time.
To this end, consider
\begin{eqnarray}
dz(t)\Bigg|\frac{\partial x(t)}{\partial z(t)}\Bigg|
&=& dz(0)
\Bigg|\frac{\partial z(t)}{\partial z(0)}\Bigg|
\Bigg|\frac{\partial x(t)}{\partial x(0)}\Bigg|
\Bigg|\frac{\partial x(0)}{\partial z(0)}\Bigg|
\Bigg|\frac{\partial z(0)}{\partial z(t)}\Bigg| \;.
\end{eqnarray}
By hypothesis $|\frac{\partial x(t)}{\partial x(0)}|=1$
since $x$ coordinates are canonical.  Hence
\begin{eqnarray}
dz(t)\Bigg|\frac{\partial x(t)}{\partial z(t)}\Bigg|
&=& dz(0)
\Bigg|\frac{\partial z(t)}{\partial z(0)}\Bigg|
\Bigg|\frac{\partial x(0)}{\partial z(0)}\Bigg|
\Bigg|\frac{\partial z(0)}{\partial z(t)}\Bigg| \;.
\end{eqnarray}
Noticing that $|\frac{\partial z(t)}{\partial z(0)}|
|\frac{\partial z(0)}{\partial z(t)}|=1$, one gets the result
\begin{eqnarray}
dz(t)\Bigg|\frac{\partial x(t)}{\partial z(t)}\Bigg|
&=&
dz(0)\Bigg|\frac{\partial x(0)}{\partial z(0)}\Bigg|
\;.  \label{eq:invmeasure1}
\end{eqnarray}
Equation~(\ref{eq:invmeasure1}) shows that the Jacobian of the non-canonical
transformation of coordinates automatically provides 
the invariant measure in phase space.
It is worth to remark that no special property of the dynamics
was used in the derivation and, in particular, no mention
has been made of phase space compressibility.

Tuckerman et al~\cite{tuckerman} defined an invariant measure for systems
with a  non-zero compressibility in phase space.
Such systems may be either non-canonical or non-Hamiltonian.
In order to arrive at the definition of the invariant measure,
they considered a Jacobian associated to phase space flow
\begin{eqnarray}
{\cal J}_t(t,t_0)&=&\Bigg|\frac{\partial z(t)}{\partial z(t_0)}\Bigg|
\;.
\end{eqnarray}
This Jacobian obeys the equation of motion
\begin{equation}
\frac{d}{dt}{\cal J}_t=\kappa {\cal J}_t\;,
\end{equation}
which can be integrated to yield
\begin{eqnarray}
J_t(t,t_0)&=&e^{\int_{t_0}^t\kappa(t')dt'}
=e^{w(t)-w(t_0)}\nonumber\\
&=&e^{w(t)}e^{-w(t_0)}\;,
\end{eqnarray}
where $w$ is the primitive function of the compressibility $\kappa$.
Then the invariant measure can be defined as~\cite{tuckerman}
\begin{equation}
dz(t)e^{-w(t)}=dz(t_0)e^{-w(t_0)}\;.  \label{measure_tuck}
\end{equation}

For non-canonical systems with compressibility
\begin{eqnarray}
{\cal J}_t(t,t_0)
&=&\Bigg| \frac{\partial z(t)}{\partial x(t)} \Bigg|
\Bigg|\frac{\partial x(t_0)}{\partial z(t_0)}\Bigg| \;,
\end{eqnarray}
where it has been used  $|\partial x(t)/\partial x(t_0)|=1$, which is
true since the $x$ coordinates are canonical by hypothesis. 
In this case, one is led to the identification
\begin{eqnarray}
e^{w(t)}&=&\Bigg| \frac{\partial z(t)}{\partial x(t)} \Bigg|
\;, \\
e^{-w(t_0)}&=&
\Bigg|\frac{\partial x(t_0)}{\partial z(t_0)}\Bigg| \;,
\label{eq:e-w}
\end{eqnarray}
so that the invariant measure proposed by Tuckerman et al,
Eq.~(\ref{measure_tuck}), and that given in Eq.~(\ref{eq:invmeasure1})
agrees.
However, if a particular non-canonical system (${\cal J}\neq 1$ )
has no compressibility, one finds ${\cal J}_t=1$ 
so that the invariant measure given in Eq.~(\ref{measure_tuck})
cannot be used. Instead, the invariant measure given
in Eq.~(\ref{eq:invmeasure1}) is still correct.

Proper consideration of the invariant measure is critical
when defining the entropy 
functional~\cite{tuckerman,andrey,ramshaw}.
For a Hamiltonian system, Gibb's entropy $S$ is given 
by~\cite{balescu}
\begin{equation}
S=-k_B\int dxf(x)\ln f(x)\;.
\end{equation}
Performing a non-canonical transformation of variables
one must take into account the Jacobian
so that in the new coordinates the entropy is given by:
\begin{equation}
S=-k_B\int dz\rho(z) \ln f'(z)\;,
\label{eq:entropy}
\end{equation}
where $\rho(z)$ is defined in Eq.~(\ref{rhoz}).
If the non-canonical system has a non-zero compressibility
$\rho(z)=e^{-w} f'(z)$ and Eq.~(\ref{eq:entropy})
would be equal to that proposed by Tuckerman.
Equation~(\ref{eq:entropy}), with $\rho(z)$ defined in Eq.~(\ref{rhoz}),
is more general since it also describes non-canonical systems
with zero compressibility.

\section{Conclusions}\label{sec:conclusion}

A formalism for Hamiltonian systems in non-canonical coordinates
has been presented. 
Non-canonical equations of motion have been derived directly from a variational 
principle written in symplectic form.
It has been shown how a certain class of
non-Hamiltonian equations of motion may be derived by means
of coordinate-dependent scalings of time.
Finally,  the invariant measure of phase space
and the covariant definition of the entropy functional
have been obtained from the Jacobian of non-canonical transformations of
coordinates.
There may be cases where the dynamical compressibility is zero.
However, non-canonical coordinates have in general a non-trivial
Jacobian that must be considered in order to
define the invariant measure.

\end{document}